# Revisiting stress-corrosion cracking and hydrogen embrittlement in 7xxx-Al alloys at the near-atomic-scale

*Martí López Freixes[1], Xuyang Zhou[1], Huan Zhao[1], Hélène Godin[2], Lionel Peguet[2], Timothy Warner[2], Baptiste Gault[1,3]\**

[1] *Max-Planck-Institut für Eisenforschung GmbH, Max-Planck-Str. 1, 40237 Düsseldorf, Germany*

[2] *C-TEC, Constellium Technology Center, Parc Economique Centr'alp, CS 10027, Voreppe, 38341 cedex, France*

[3] *Department of Materials, Royal School of Mines, Imperial College London, Prince Consort Road, London SW7 2BP, UK*

*\* corresponding authors. E-mail addresses: b.gault@mpie.de*

## Abstract

Hydrogen embrittlement (HE) affects all major high-strength structural materials and as such is a major impediment to lightweighting e.g. vehicles and help reduce carbon-emissions and reach *net-zero*. The high-strength 7xxx series aluminium alloys can fulfil the need for light, high strength materials, and are already extensively used in aerospace for weight reduction purposes. However, depending on the thermomechanical and loading state, these alloys can be sensitive to stress-corrosion cracking (SCC) through anodic dissolution and hydrogen embrittlement. Here, we study at the near-atomic-scale the intra- and inter-granular microstructure ahead and in the wake of a propagating SCC crack. Moving away from model cases not strictly relevant to application, we performed an industry-standard test on an engineering Al-7XXX alloy. H is found segregated to planar arrays of dislocations and to grain boundaries that we can associate to the combined effects of hydrogen-enhanced localized plasticity (HELP) and hydrogen-enhanced decohesion (HEDE) mechanisms. We report on a Mg-rich H-rich amorphous oxide on the corroded crack surface and evidence of Mg-related diffusional processes leading to dissolution of the strengthening $\eta$-*phase* precipitates ahead of the crack. We show ingress of up to 1 at% O, i.e. well above the solubility limit of O in Al, near the oxide-metal interface, while no increased level of H is found in the matrix. We provide an array of discussion points relative to the interplay of structural defects, transport of solutes, thereby changing the resistance against crack propagation, which have been overlooked across the SCC literature and prevent accurate service life predictions.

# 1 Introduction

The global need to achieve net-zero carbon emission reinforces the need for light and high strength materials for e.g. light-weighting vehicles and developing the infrastructure for a hydrogen economy in e.g. the commercial aircraft[1] or ironmaking[2] sectors. The 7xxx series Al-alloys are already extensively used for weight reduction purposes in aerospace[3] and are poised to see their usage widen. Cu-rich 7xxx series Al-alloys have complex microstructures, including $\eta'/\eta$-phase (Mg(Zn, Cu, Al)$_2$) hardening precipitates, dispersoids such as Al$_3$Zr, and coarse intermetallic particles such as Al$_7$Cu$_2$Fe and Mg$_2$Si [4]. These alloys are strengthened through precipitation of the solute-rich $\eta$-phase in the grain interior and at the grain boundary (GB). However 7xxx Al-alloys can be susceptible to hydrogen embrittlement (HE), particularly through stress-corrosion cracking (SCC) [5–9].

SCC is driven by both anodic dissolution and HE, but their relative contributions to SCC depend on the environment but also the alloy compositions, plate thicknesses and tempers [5,6,10–12]. Academic studies often focus on the effects of H on mechanical properties or its microstructural trapping[13–18] under conditions far from those encountered in operation, making them difficult to directly translate into practice. Despite extensive research on SCC and HE of these alloys[5,9,11,19], a holistic, mechanistic understanding of these complex processes is still missing and this hinders the development of materials design strategy to overcome these crucial issues.

Atomic H is produced at crack tips as a consequence of corrosion [13,20–22] mostly arising from differences in potential between the $\eta$-phase, Al-rich matrix [23] and the GB region, and is hence indissociable from SCC. HE has been associated to hydrogen-enhanced localized plasticity (HELP) and hydrogen-enhanced decohesion (HEDE) of GBs amongst others[9,14,24]. The HELP mechanism is based on an observed increase in dislocation motion and slip localization in the presence of H in solid solution[24], whereas HEDE proposes that the presence of H reduces the cohesive energy of interfaces, thereby reducing the fracture work [14]. A recent article proposed that HEDE is caused by the joint effect of H and Mg segregated to GBs[16].

Artificial ageing, used to optimize properties, also alters the composition of the precipitates and their susceptibility to corrosion [25,26] and SCC [7,8,11,12,27,28]. Overageing, introduces more Cu in the $\eta$-phase and increases its electrochemical potential and thus reduces the potential difference with the matrix and diminishes H generation at the cathodic sites, which has been associated with a reduced susceptibility to SCC [7,29,30]. Cracks caused by SCC propagate primarily intergranularly and crack arrest markings (CAM) observed on fracture surfaces in chloride solutions[6] and humid air[31] indicate discontinuous cracking, which is inconsistent with anodic dissolution processes being rate-controlling[32] as sometimes proposed[5,11,12].

The presence of CAMs suggests that all cracking-related phenomena occur locally, just ahead of the crack tip. In addition, the oxide produced by corrosion, and its relative stability determine the H production rate, its ingress and trapping are microstructure dependent. Yet, we lack spatially resolved H compositional data during SCC and there is very limited information reported on the composition and structure of the corroded layer formed at the tip of a propagating stress-corrosion crack in Al-alloys. These knowledge gaps hinder the determination of the active mechanisms during SCC.

Here, we study the microstructural and microchemical changes caused by a propagating stress-corrosion crack at the nanoscale by using transmission-electron microscopy (TEM) and atom probe tomography (APT). APT has a high chemical and spatial resolution, and provides accurate compositional information [33]. It has successfully been used to map trapped H in steels [34–36], including at crystalline defects [37], as well as in Ti-alloys [38], in a 7xxx Al-alloy [16], and to analyse crack tip oxides formed during SCC in stainless steels [39–41]. We focus on the direct vicinity of cracks and in regions ahead of the main crack, especially grain boundaries along which the crack is expected to grow. We observe the segregation of H to the GB ahead of the crack and on linear features usually attributed to dislocations. The corroded crack surface is a Mg-rich, H-rich chlorinated amorphous oxide. For clarity, we refer to it simply as oxide throughout. The composition of the PFZ and of precipitates near and ahead of the crack is vastly modified, and the matrix near the metal-oxide interface contains up to over 1 at% O, i.e. well above the solubility limit of O in Al, while no increased level of H is measured. We discuss our findings in perspective with the published literature and provide an array of discussion points relative to the interplay of structural defects and the transport of solutes from the matrix, possibly assisting with the dissolution of the strengthening phases, thereby changing the resistance against crack propagation.

## 2 Results

### 2.1 Sample characterization far from the crack

APT analyses of the commercial aluminium alloy 7449-T7651 from the GB and the grain interior away from the crack including, $\eta$-phase precipitates, matrix and PFZ are displayed in Figure S1. Features across multiple datasets show spread in their compositions, as plotted in Figure S1b-c. Mg and Zn levels within the $\eta$-phase precipitates from the grain interior and the near GB region display a linear relationship, with Mg ranging from 15 to 32 at.%, and Zn from 25 to 54 at.%. In turn, matrix and PFZ solute compositions exhibit an array of possible values. The mean Cu value for each of the features of interest are shown in Figure S1b-c. The average composition values of the matrix, PFZ and the $\eta$-phase GBP are reported in Table 1.

### 2.2 Effect of H on deformation behaviour

We performed an industry standard DCB crack growth test to study the SCC behaviour of the alloy 7449-T7651. We investigated the crack tip region using (S)TEM. The crack is intergranular and displays void-like features (Figure 1a), suggested by Lynch [9] to be one of the crack propagating mechanisms. Energy-dispersive X-ray spectroscopy (EDS) confirms the presence of an oxide at the crack and within the voids, and the matrix adjacent to the crack oxide is devoid of precipitation (Figure S3 and Figure S4). The selected area electron diffraction pattern from the corroded crack indicates that the oxide is amorphous (Figure S5). Near the void-like features, we observed an array of dislocations (Figure 1b), indicative of plasticity, sitting on (111) planes, as confirmed by dark field TEM (inset Figure 1b).

APT analysis was performed approx. 4 μm ahead of a crack tip (Figure 1c). The dataset contains a GB with $\eta/\eta'$-phase precipitates as well as $Mg_2Si$ particles. The red set of iso-surfaces delineates regions of high H composition that form a planar array of elongated features (Figure 1d) typically associated with segregation to dislocations [42–44]. Based on atom probe crystallographic analysis, as outlined in ref. [45], we confirmed that these features lie on a (111) plane, as expected for dislocations in the face-centred cubic Al-matrix, thereby supporting the hypothesis. The segregated H reaches locally near 20 at. %, along with approx. 4 at. % Si. H is also found segregated to the grain boundary and trapped within the $Mg_2Si$ particle sitting at the boundary. This is in line with recent reports of H segregation to grain boundaries in a 7xxx Al-alloy [16]. The observed segregation to the dislocations could thus be caused by the shearing through $Mg_2Si$ particles, and both Si and H being dragged by, or subject to pipe diffusion along the dislocation connecting regions of different chemical potential [46–48]. H measurements by APT are notoriously challenging, because of the presence of H from the residual gas of the ultrahigh vacuum chamber [49]. The reliability of our measurements is further discussed below.

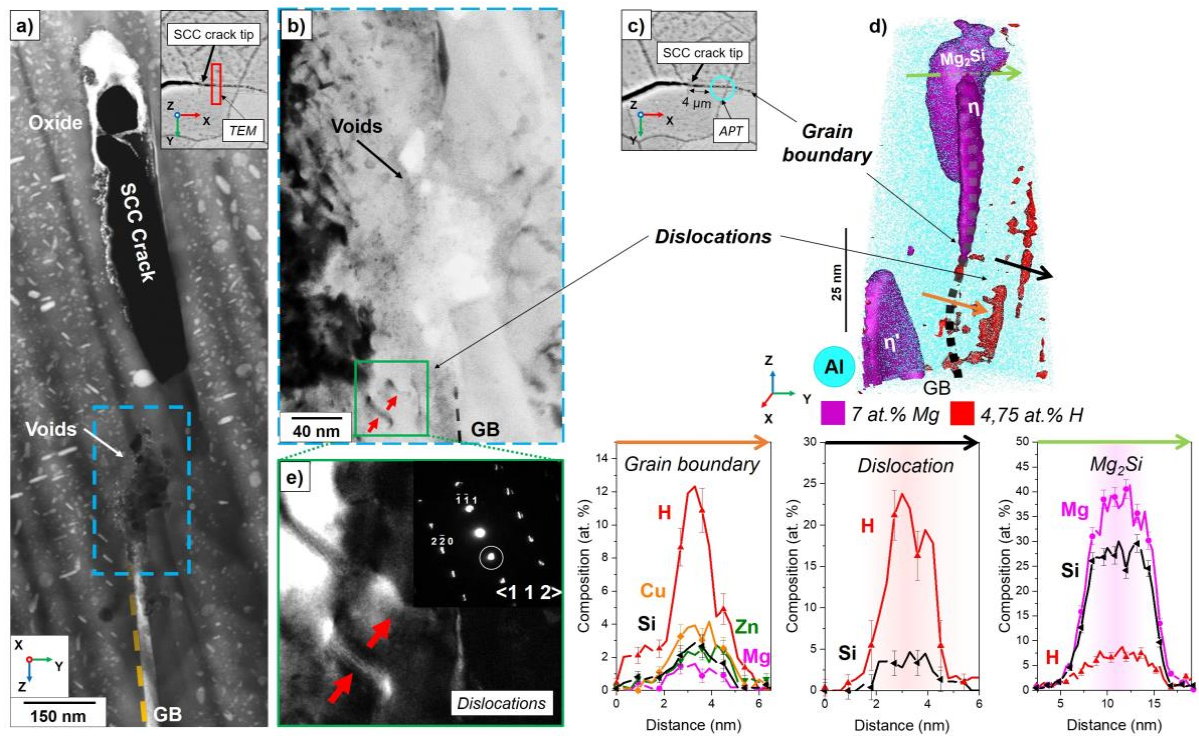

*Figure 1. H effects on deformation behaviour near the SCC crack. a) STEM image of the crack tip region revealing the presence of oxide and void-like structures. b) TEM BF image showing dislocations near the crack. c) Schematic of the APT sample location with respect to the crack tip. d) APT reconstruction of the GB 4μm ahead of the crack showing the presence of dislocations with H and Si segregation. 1D Composition profiles across the GB, the dislocation and the $Mg_2Si$ particle, respectively measured along the arrows shown in the APT reconstruction. The error bars correspond to the standard deviation within each of the bins in the profile. e) DF image corresponding to the (111) diffraction spot of the dislocations imaged in b). Note that the SEM micrograph in c) and inset in a) is indicative of the analysis location. The results displayed herein were obtained from different cracks tips.*

## 2.3 On the crack oxide

We targeted the tip of an SCC crack for specimen preparation along the main propagating crack in the DCB sample (Figure 2a). Analysis of the desorption pattern (Figure S6) formed through the APT dataset (Figure 2b) provided sufficient crystallographic information to determine the existence of two grains and thus of a GB within the reconstruction. We therefore assumed that we captured the tip of the actual propagating crack, as SCC cracks are known to propagate almost exclusively following an intergranular mode in Al-alloys [5]. However, we could not locate the GB within the dataset, as no clear evidence of segregated solutes could be found, and there was a microfracture [50] during the data acquisition that could lead to a loss of information specifically in the GB region.

An iso-composition surface with an arbitrary threshold of 12 at.% O highlights the oxide-matrix interface, which exhibits a very complex, rough morphology (Figure 2b). Figure 2c plots the atomic ratios through the interface, along the suspected crack propagation direction. A two-layer structure appears, with first a ~5 nm zone enriched in Mg and O, followed by a region with increased Zn and lower in O and Mg. O/H increase again further ahead, but Mg and Zn do not show increased levels. The formation of an Mg-rich oxide film in aluminium alloys was suggested by Green et. al. [51] and reported experimentally [20,22]. The oxide film is also found to be hydrated, which is in line with previous observations [5,20,21]. Levels of Cl of up to 6 at%, not included in the profiles for clarity, are also incorporated in the oxide film, as expected [6].

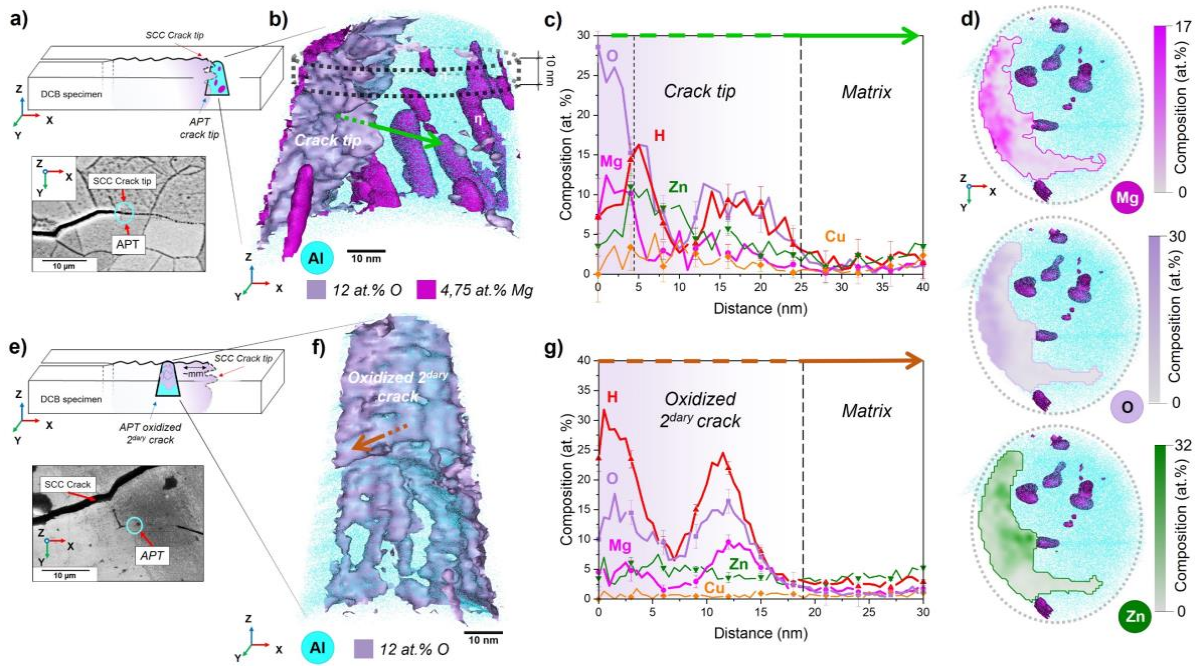

*Figure 2. Crack corrosion behaviour during SCC. a) Schematic of a DCB sample showing the APT crack tip sample location and orientation. b) APT reconstruction of the crack tip of an SCC crack. c) 1D composition profile across the oxide-matrix interface showing the oxide composition, measured within a 10 nm (ø) cylinder along the arrow in b). d) 2D composition maps, obtained from a 10 nm slice, showing the 2D elemental distribution of Mg, O and Zn in the oxide within the dotted-line slice shown in b). e) Schematic of a DCB sample showing the APT secondary crack sample location and orientation. f) APT reconstruction of an oxidized secondary crack displaying the complex oxide morphology and absence of precipitation. g) 1D composition profile across the oxide-matrix interface, measured within a 10 nm (ø) cylinder along the arrow in f). The error bars correspond to the standard deviation within each of the bins in the composition profiles shown.*

Two-dimensional compositional maps from the top of the reconstructed APT dataset as indicated in Figure 2d, and displayed top-down, highlight the inhomogeneity of the oxide and complement Figure 2c. The enrichment of Mg and O is coincident, while Zn and Cu (Figure S7) appear to be accumulated at the oxide-metal interface. This suggests that Mg is being preferentially oxidized and that dealloying is occurring during corrosion of the freshly cracked surface. An Mg-rich and Zn-poor oxide and a Zn-rich, Mg-poor adjacent region is consistent with previous reports on oxidized $\eta$-phase precipitates [30,52,53].

We then performed APT analyses several mm in the wake of the crack tip, along an oxidized secondary crack (Figure 2e), in order to reveal the solute distribution long after the crack has passed. The tomographic reconstruction in Figure 2f also evidences a complex morphology of the interface between the metallic matrix and the oxidised crack. A composition profile perpendicular to the oxide/metal interface (Figure 2g) shows the highly inhomogeneous oxide composition, with H levels reaching up to 30 at.% in the film, and with O/H peaks coinciding with Mg peaks. Conversely, Zn and Cu display no partitioning. The adjacent matrix is devoid of precipitates, as observed also by STEM-EDS (Figure 1 and Figure S3). This motivated an analysis of the composition of the $\eta$-phase precipitates around the crack tip (Figure 3a) that reveals a lower solute content than expected. The Mg content decreases down to 13 at.% and Zn levels to 20 at.% within the precipitates, on average leaner in solute compared to the reference composition reported in Figure S1**Error! Reference source not found.**b. The behaviour of Cu is similar to Mg and Zn, with Cu levels going from 7,3 at.% in the reference measurements (**Error! Reference source not found.**Figure S1b) to 2,7 at.% in the $\eta$-phase precipitates around the crack tip (Figure 3a).

The precipitates' dissolution leads to a significant increase in the matrix composition (Figure 3b), up to tenfold across different locations around the crack tip with respect to the undeformed reference (Figure S1a). This is also observed in the matrix adjacent to the oxidized secondary crack (Figure 3b and Table 1), where the composition is close to the alloy's nominal composition (Table 2), except for Mg, which has leached to the oxide layer. The matrix composition in the vicinity of the crack tip bridges the gap between the near-crack region and the reference data, hinting at a transient precipitate dissolution process (Figure 3b). The variation in the matrix compositions is notably higher than in the reference measurements and no correlation between the solute enrichment and the distance to the captured oxides could be established. We also quantified the O dissolved in the matrix which, to a first approximation, scales with the increasing matrix solute content, as reported by the colour change of the symbols in Figure 3b, and is far above the expected content [54].

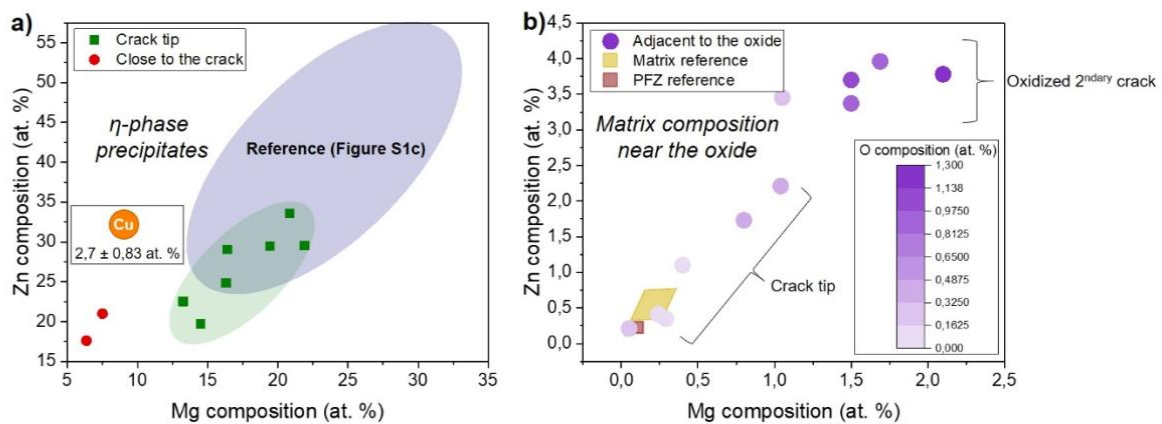

Figure 3. Matrix and η-phase precipitate compositional differences near the oxide a) η-phase precipitate composition around the crack tip and very close to the crack compared with reference values from Figure S1c. The composition of η-phase precipitates captured close to the crack, within 1 µm, in another dataset are also added to the analysis. Cu average composition in the η-phase precipitates at the crack tip and close to the crack is also reported, also showing a marked decrease in Cu content with respect to reference values from Figure S1c. The error values reported for Cu correspond to the standard deviation of all measurements; b) matrix composition around the crack tip and adjacent to the oxidized secondary crack compared with reference values from Figure S1b. Matrix O levels are also displayed at each data point. Please refer to Figure S2b for error values for O measurements.

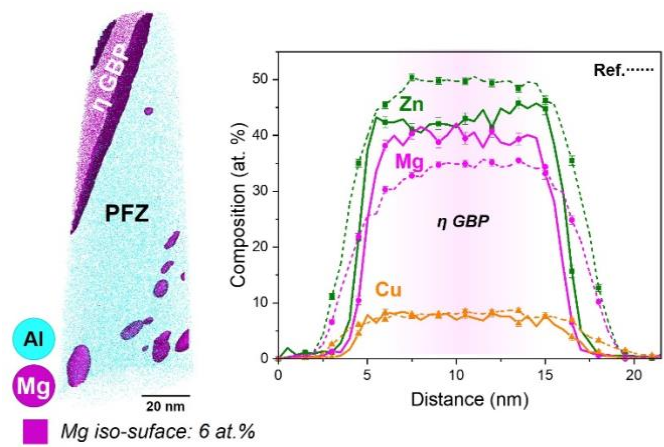

Figure 4. Grain boundary ahead of the crack tip. APT reconstruction of the grain boundary ahead of the crack, showing a η-phase grain boundary precipitate with an Mg composition increased with respect to the 33 at.% dictated by the stoichiometry of the Mg(Zn, Cu, Al)$_2$ phase. The error bars correspond to the standard deviation within each of the bins in the profile.

## 2.4 Composition fluctuations ahead of the crack

APT analyses near or at the grain boundary ahead of the crack yielded more evidence of the dynamic behaviour of the microstructure during SCC. APT analysis of a $\eta$-phase grain boundary precipitate located approx. 2 µm ahead of the crack shows an increased Mg concentration with respect to the expected stoichiometry of $Mg(Zn,Cu,Al)_2$. The analysis reveals 38 at.% Mg, 46 at.% Zn, and 7 at.% Cu on average through the whole precipitate, but with 40 at. % Mg being reached locally (Figure 4). This difference is significant based on the standard deviations for a GBP measured in the reference sample (Table 1).

*Table 1. Calculated average values from the reference data shown in Figure S1a-b. The values for the matrix solute content in the oxidized $2^{ndary}$ crack are an average of the values presented in Figure 3b. The error values reported correspond to the standard deviation of all measurements.*

|  | Al | Zn | Mg | Cu |
|---|---|---|---|---|
| Matrix oxidized $2^{ndary}$ crack | 94.13 ± 0.59 | 3.53 ± 0.34 | 1.63 ± 0.42 | 0.72 ± 0.09 |
| PFZ (Ref.) | 99.68 ± 0.16 | 0.18 ± 0.04 | 0.08 ± 0.03 | 0.06 ± 0.03 |
| Matrix (Ref.) | 98.47 ± 0.48 | 0.54 ± 0.16 | 0.19 ± 0.07 | 0.1 ± 0.07 |
| GBP (Ref.) | 8.83 ± 1.26 | 49.2 ± 0.68 | 34.4 ± 0.54 | 7.53 ± 0.33 |

# 3 Discussion

Our results provide evidence for H absorption and segregation to dislocations and GBs during SCC in 7xxx aluminium alloys. HE caused by the ingress of H ahead of the crack tip generally causes a "brittle" fracture surface appearance and CAMs [55]. The embrittling effect of H had been so far been proven indirectly[56]. In-situ TEM experiments have shown that H promotes slip planarity, enhanced dislocation glide and a decreased flow stress [57,58]. H transport by dislocations was already proposed by Albrecht et al. [59] in 1982, but had not been directly measured in aluminium alloys, especially during SCC. We provide here (Figure 1d) the first evidence of co-localisation of H and dislocations in an Al-alloy, indicative of segregation and/or localised transport. H segregating to dislocation cores and promoting slip localization and H transport forms the ground of HELP, i.e. one of the prevalent HE models [14,24,60], and our observations strongly suggest it that it is active during SCC. As samples are over 3 years old, H must have been deeply trapped by the dislocations, consistent with previous reports [15].

In addition, we present evidence of H segregating to grain boundaries ahead of the crack during SCC (Figure 1d), suggesting that HEDE could be contributing to the crack progression, reducing the grain boundary cohesion based on the mechanism proposed by Zhao et al. [16]. Since H was detected along the GB, within an $Mg_2Si$ intermetallic particle and at dislocations in the matrix near the grain boundary 4µm ahead of the crack, H diffusion through the grain boundary must have occurred during the crack propagating or as the cracked surface oxidised.

The evolutions of the microstructure and microchemistry near a propagating SCC crack had not previously reported in Al-alloy, but are consistent with previous observations in other engineering alloys[41,61,62]. We evidence a specific behaviour for Mg compared to Zn and Cu. The distribution reported is after the solutes were given ample time to equilibrate at room temperature, as the samples were only analysed approx. 3 years after the tests. We did not measure an increased H content in the Al-matrix, but an increased O content in the solute-rich sub-oxide region. O solubility in Al is reported to be close to zero [54]. Vacancy injection during cationic oxide growth could play a role, but recent investigations report negligible vacancy-O binding energy [63]. The origin of this increased O solubility is unclear and the effects on the mechanical properties such increased O levels in Al are unknown. In e.g. Ti-alloys [64],

O is a known solid solution strengthener, and O-ingress during SCC was shown to impact local phase equilibrium and cause the precipitation of an embrittling phase[62]. This could have an influence on SCC performance, and hence be relevant to the cracking process, in particular as this is occurring locally, just ahead of the crack, adjacent to the growing oxide.

Al oxide films at room temperature form by outward diffusion of $Al^{3+}$ ions, according to the Cabrera-Mott model and experimental measurements [65,66]. This requires a steady diffusion of atoms from the bulk through the immobile metal/oxide interface [66]. Here, for the first time in Al-alloys, we analysed the composition of the oxide formed at the tip and in the wake of an SCC crack in a 7xxx series Al alloy. We show that a preferential Mg oxidation is taking place during SCC, thus creating an Mg-rich and also H-rich amorphous oxide. Interestingly, the oxide composition is different to measurements by Auger spectroscopy on AA7010 in dry air or seawater, which consisted of an outer MgO layer and an inner $Al_2O_3$ [67].

The Mg partitioning to the oxide provides a driving force for Mg diffusion in the near-crack region. Diffusion from the neighbouring matrix and along grain boundaries to the Mg-depleted region is hence expected upon the onset of corrosion, thereby disturbing the local equilibria and causing dissolution of precipitates to feed the progressing crack and growing Mg-rich oxide. The inhomogeneous solute supply to the crack during oxidation results in compositional fluctuations within the oxide (Figure 2g). Precipitate dissolution and solute transport to an oxidizing surface, possibly accelerated by diffusion along defects, occurs readily in the absence of stress [20,68]. This was referred to as the "reservoir effect" by Young and Gleeson [69] and has been observed in other systems like Cr-containing Fe-based alloys [70] for instance. Precipitates can then be understood as potential solute sinks that are thermodynamically trapped as a result of the heat treatment.

Importantly, the dissolution of $\eta$-phase precipitates would eliminate obstacles to dislocation glide ahead of the crack and release previously pinned dislocations, thereby altering the mechanical properties of the material ahead of the crack. Dislocations could provide fast diffusion pathways [47] for the solutes rejected to the matrix to support the growth of the oxide. In the matrix composition values adjacent to the oxidized secondary crack (Figure 3b), Mg displays higher relative variation than Zn, which could indicate enhanced Mg diffusion to the crack along defects. However, none of the main alloying elements was observed at the dislocations imaged here (Figure 1d), so their role in assisting diffusion cannot be assessed from the current data. H segregation to dislocation cores increases planar slip, and thus increases strain localization [18]. If precipitates are being sheared by dislocations, this can accelerate their dissolution [71–73].

Since crack propagation takes place mostly along or near to the GBs, these could also act as a fast solute diffusion pathway to drive solutes towards the crack. The increased Mg composition found in an $\eta$-phase precipitate suggests that accelerated solute diffusion is also occurring along the grain boundary (Figure 4). Alani [68] found large Zn-rich precipitation on the GB after corrosion occurred on the adjacent surface, which would be consistent with the diffusion-induced dissolution and re-precipitation at the grain boundary. They attributed the anomalous GB precipitation to an enhanced GB diffusion caused by the exposure to humid air at a relatively high temperature .

As a result of the Mg preferential consumption during the growth of the oxide, a subsurface region of differing composition forms [69,74], rich in Zn and Cu, and depleted in Mg. We did not try to perform a precise diffusion length analysis, as the actual depth of the subsurface layer cannot be readily obtained from the APT data presented here, which could potentially lead to error in discerning the actual mechanisms contributing to diffusion in this region. Similar APT observations of substantial compositional changes and associated dissolutions in oxidised and stressed regions in superalloys had previously been reported [61,75]. In Al-alloys, if corrosion is occurring at a high enough temperature, Alani and Schwarzenböck et al. [20,68] show Zn-rich precipitates forming in this solute-rich subsurface region, following oxidation in high humidity environments at 70°C and 80°C respectively. These observations are consistent with Mg diffusion to the oxide, precipitate dissolution and re-precipitation, similarly to what they observed in the matrix. However, it appears dependent on alloy composition, i.e. the Zn/Mg ratio, as Schwarzenböck et al. [20] only found anomalous subsurface precipitation in the alloy with the higher Zn/Mg ratio (5,2). It should be noted that the materials used in ref. [20] were industrially sourced, so differences can also arise from differing ageing practices, as a same temper (e.g T7651) will correspond to different ageing times and temperatures for different alloy compositions, including because of through-thickness microstructural and compositional variations that can affect the oxidation behaviour and H uptake.

H and O quantification by APT are known challenges, due to the presence of residual gases in the analysis chamber, ingress during specimen preparation and surface contamination during specimen transport and analysis [49,76–78]. These spurious species can obscure the signal from H or O originating from the specimen itself [49,76]. Higher electric fields during the analysis tend to improve the reliability of the measurements [38,49,76]. To assess the validity of the H/O quantification, the strength of the electric field during the experiments was determined, which can be estimated by using the charge-state ratio of Al ($N(Al^{2+})/(Al^+)$) as a proxy [79,80]. Figure S2a shows the H concentration at different features reported herein, plotted against $N(Al^{2+})/(Al^+)$. H measurements from the matrix across a dozen APT dataset obtained far from the crack were included as reference. The H levels found in the crack, the GB, the dislocation and the $Mg_2Si$ particle represent a marked increase with respect to the reference values at comparable field strengths, and with specimens prepared under the same conditions. Similarly, O values in the matrix adjacent to the oxide are plotted against the field strength in Figure S2b. It is thus clear that the increased H- and some of the O-levels reported are not caused by a local drop in the electric field strength increasing residual gas adsorption from the chamber [76], and originate from a difference in the materials' composition arising from the SCC or the processing [81].

In summary, we have studied the consequences on the microchemistry and microstructure of the propagation of a stress-corrosion crack in a 7xxx Al-alloy. The detection of aligned sets of dislocations carrying high content of H suggest strain localisation that can also assist both with precipitate dissolution and with crack propagation, according to the HELP mechanism [24]. H is also found segregated to GBs, which can reduce its cohesion and promote cracking, as stated by the HEDE mechanism [14,16]. In addition, we report on a hydrated, Mg-rich, non-stoichiometric Al-oxide. The growth of the oxide drives leaching of Mg from the alloy, and this localised dealloying in the vicinity of the crack produces a strong redistribution of the solutes, leading to precipitate dissolution. Dislocations, along with grain boundaries, also offer a path for accelerated diffusion of solutes towards the growing oxide and hence the dealloying process.

We cannot yet provide a clear answer as to whether the reported precipitate dissolution process plays a direct role in the cracking mechanism by modifying the local resistance against further crack propagation, or if it occurs concomitantly, accelerating H production and absorption by supplying more reactive solutes to the oxidizing surface. Should this be found to contribute to SCC, it would be consistent with the empirical evidence that overageing reduces crack growth rates, since it increases precipitate size and pulls solutes out of the matrix, thus rendering the oxidation, precipitate dissolution and solute redistribution processes all slower. We also reported the presence of a high concentration of O and of the main solutes in the matrix in the vicinity of the propagating crack. The possible influence of this addition of O on the local strength and other material properties remains unclear.

It is also a possibility that some of the features observed have occurred post-mortem, as the sample was close to 3 years old at the time of analysis. However, the authors could not rationalise this hypothesis based on the current available literature. We aim to perform further measurements on a different set of samples to confirm or refute the results presented herein. Yet these observations raise numerous questions that should motivate further research, in particular if the microstructure evolves and solute redistributes during SCC of 7xxx Al-alloys in humid air. These insights will be precious to help refine existing models and improve their predictive capabilities.

**Acknowledgements**

We thank U. Tezins, A. Sturm, M. Nellessen, C. Broß, and K. Angenendt for their technical support at the FIB/APT/SEM facilities at MPIE. We are grateful for fruitful discussions and help with experiments from Diana Koschel, Annabelle Rossetto, Marcel Glienke and Damien Connétable. B.G. acknowledge the financial support from the ERC-CoG-SHINE-771602. X.Z. is grateful for funding from the Alexander von Humboldt Stiftung.

**Method**

We conducted a double cantilever beam (DCB) test on specimens machined in the S-L orientation from the mid-thickness of 75 mm 7449-T7651 plate supplied by Constellium. Before the test, the surface was thoroughly wiped with acetone and subjected to ultrasound cleaning. Testing was carried out according to ASTM G168-17 [82] at room temperature in laboratory air. Pre-cracking was done manually by "pop-in" and droplets of 3.5% NaCl solution were added twice a day in the slit during weekdays. The DCB test was carried out in Constellium' laboratory, nearly 3 years before the microstructural characterization was performed.

Table 2. Nominal compositions for AA7449 (in at%, calculated from wt% nominals [83])

| Al   | Zn   | Mg   | Cu  | Mn  | Si   | Fe   | Zr+Ti |
|------|------|------|-----|-----|------|------|-------|
| Bal. | 3.55 | 2.65 | 0.8 | 0.1 | 0.06 | 0.05 | 0.06  |

TEM and APT specimens were prepared from the crack tip and the grain boundary ahead using a FEI Helios Xe-Plasma focused ion beam (PFIB), as Ga is known to cause GB embrittlement and quantification issues in aluminium [84]. The site-specific preparation procedure used for APT specimen preparation is outlined in Ref. [85]. APT analyses were performed on a Cameca Instrument Inc. Local Electrode Atom Probe (LEAP) 5000 XR (reflectron fitted) in voltage-pulsing mode, with a 20% pulse fraction at a pulsing rate of 125 kHz, with the specimen at a base temperature of 75, and with 4 ions detected per 1000 pulses on average. The non-oxide containing specimens were acquired at 40 K, 0.5% detection rate, 20% pulse fraction and 200 kHz pulse rate. APT reconstruction and analysis were performed using AP Suite 6.1 and reconstructions were calibrated using the procedure described in ref. [86]. (S)TEM analysis was carried out in a JEOL JEM-2200FS operated at 200kV.

**Data availability**

The datasets generated during and/or analysed during the current study are available from the corresponding author on reasonable request

**Author contributions**

Martí López Freixes: APT analyses and writing

Xuyang Zhou: (S)TEM microscopy, reviewing

Huan Zhao: Academic discussion, reviewing.

Hélène Godin: Reviewing.

Lionel Peguet: Academic discussion, reviewing.

Timothy Warner: Academic discussion, reviewing.

Baptiste Gault: Supervision of research, academic discussion, writing, reviewing.

**Declaration of competing interest**

The authors declare no competing interests.

**Materials & correspondence**

Correpondence and materials request should be addressed to B.G.

86. Gault, B. *et al.* Advances in the calibration of atom probe tomographic reconstruction. *J. Appl. Phys.* **105**, 034913 (2009).

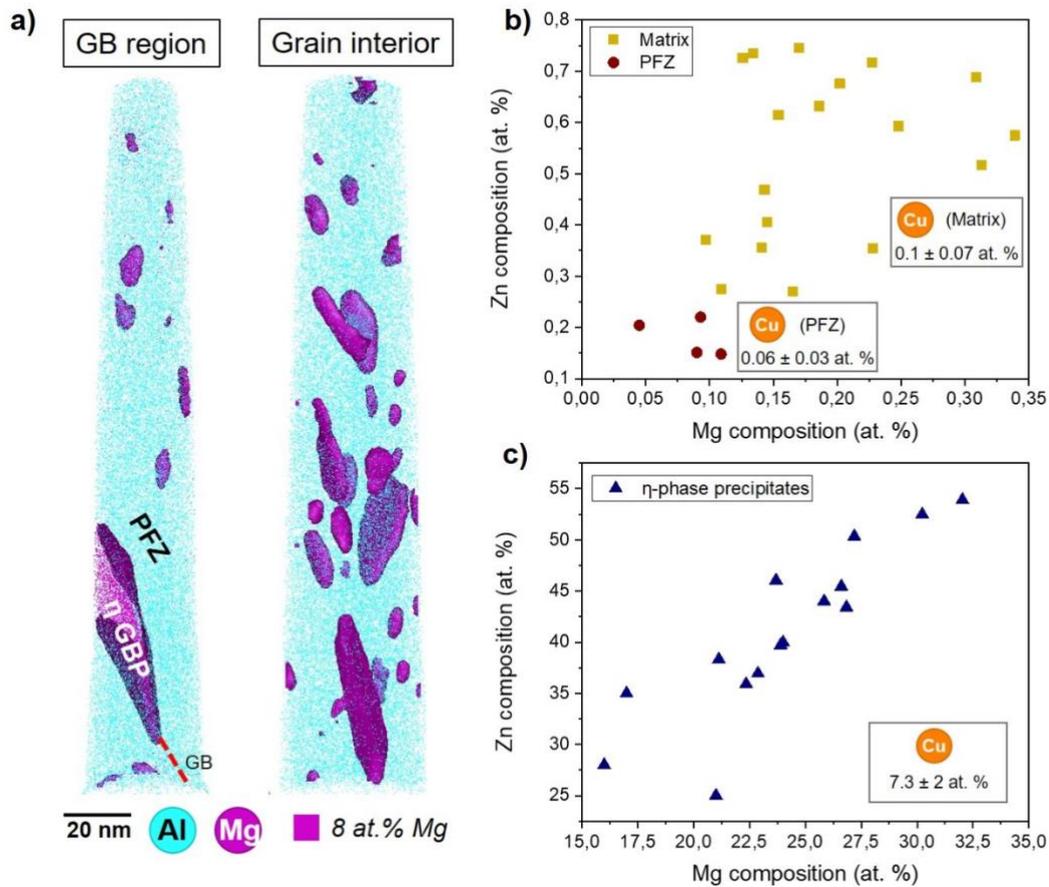

*Figure S1.* S*ample characterization, APT reconstructions of the grain boundary and the grain interior away from the crack. a) Solute matrix composition from 4 datasets and PFZ reference compositions from the dataset shown above; b) η-phase precipitates reference compositions taken from the grain interior from 3 datasets and the near GB region from the dataset shown above. Only plate shaped precipitates were analysed. The error values reported for Cu correspond to the standard deviation of all measurements.*

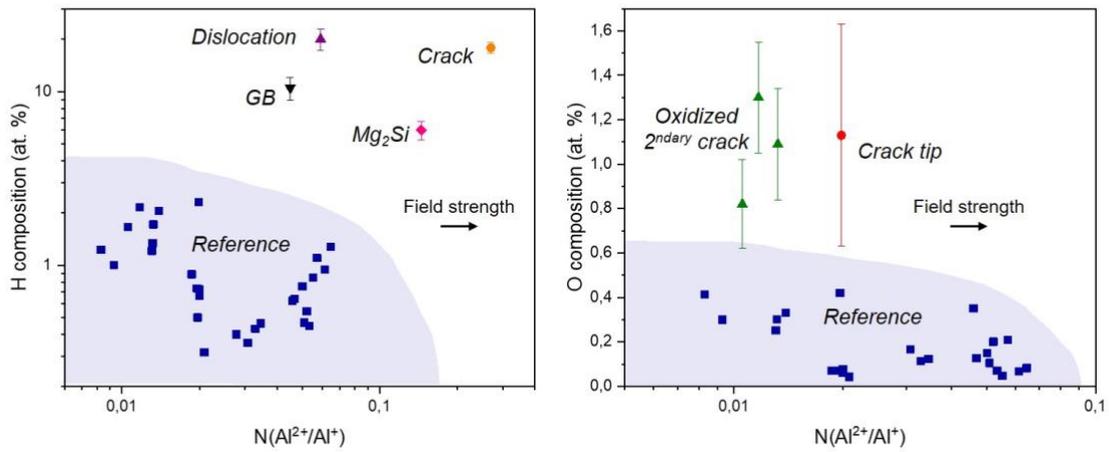

*Figure S2. Reliabilty of H/O measurements in APT. a) Hydrogen composition within the oxide in the secondary crack (Figure 2f) and at the dislocation (**Error! Reference source not found.**d) plotted against field strength during the APT experiments, showing an increase with respect to reference values at similar field strengths. b) Oxygen composition in the matrix adjacent to the crack tip (Figure 2b) and the oxidized 2$^{ndary}$ crack (Figure 2f), also showing increased levels. Values from the matrix from several datasets are included as reference.*

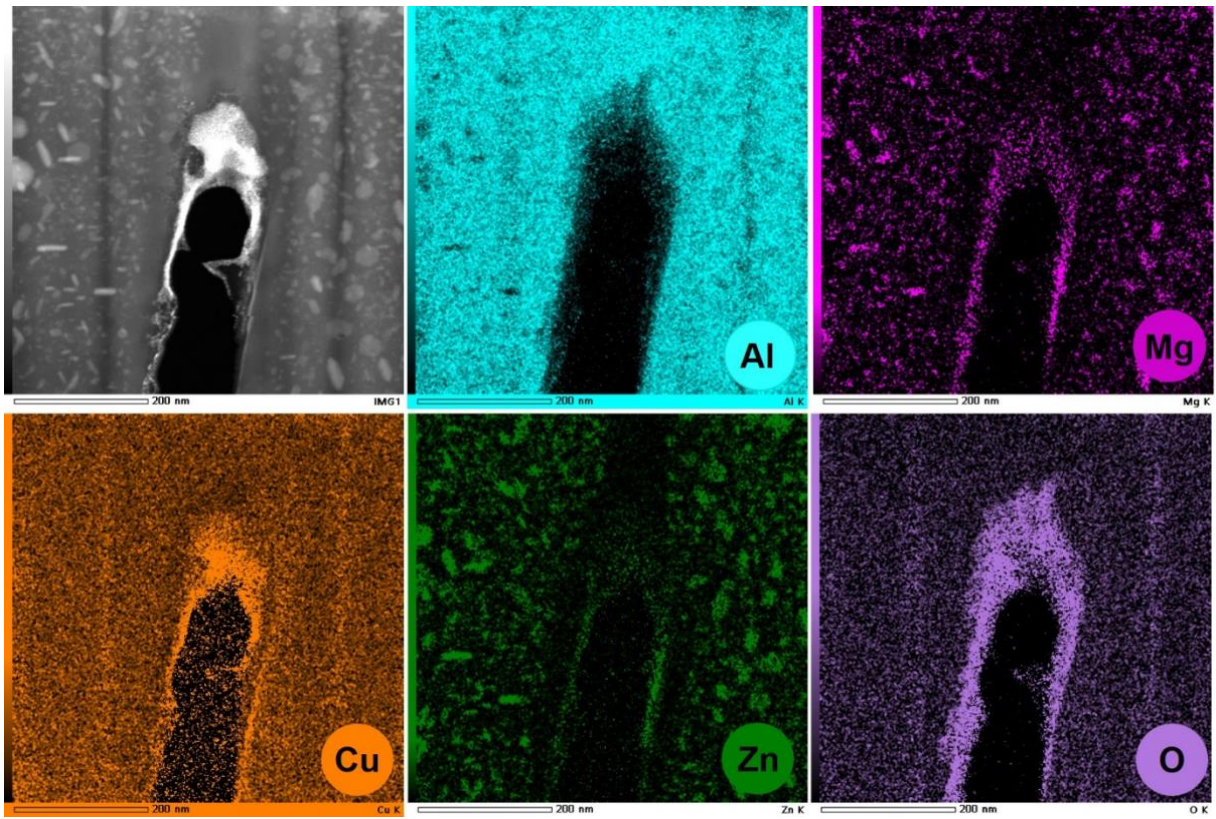

*Figure S3. EDS maps of the SCC crack shown in Figure 1a*

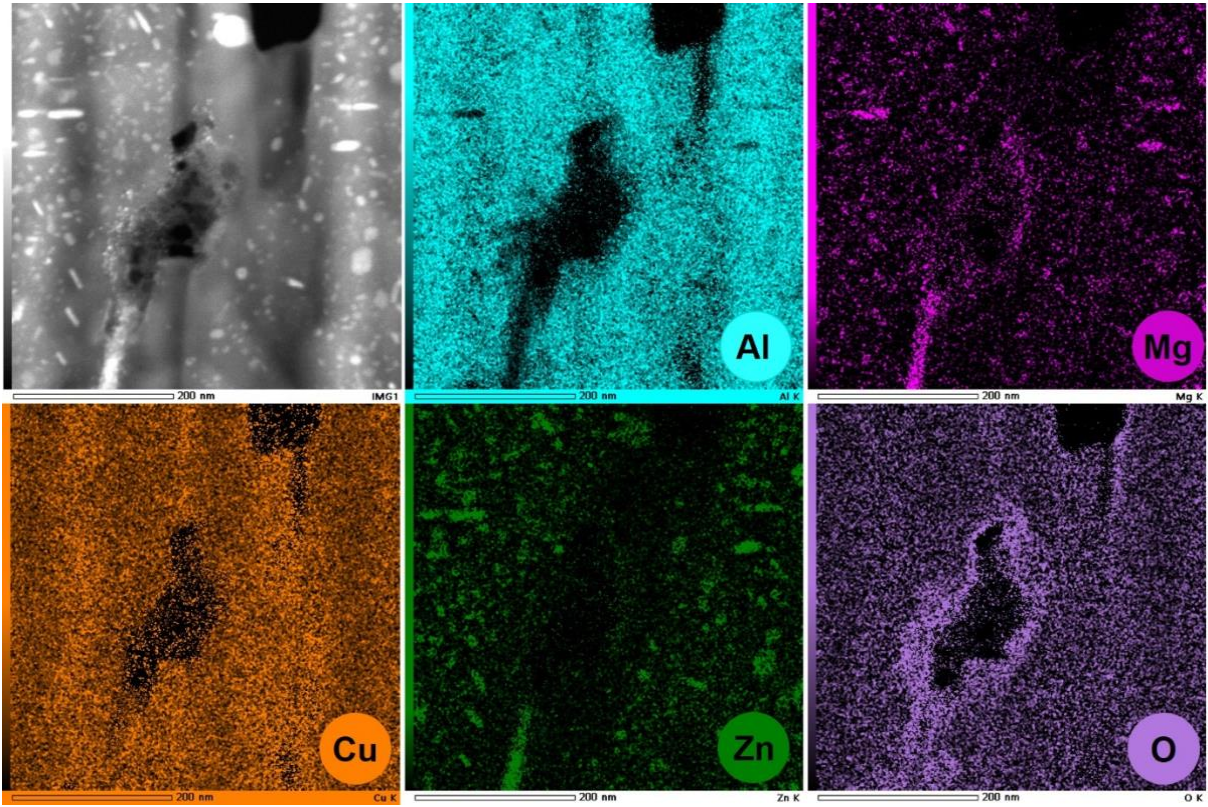

*Figure S4. EDS maps of the void-like features shown in Figure 1a*

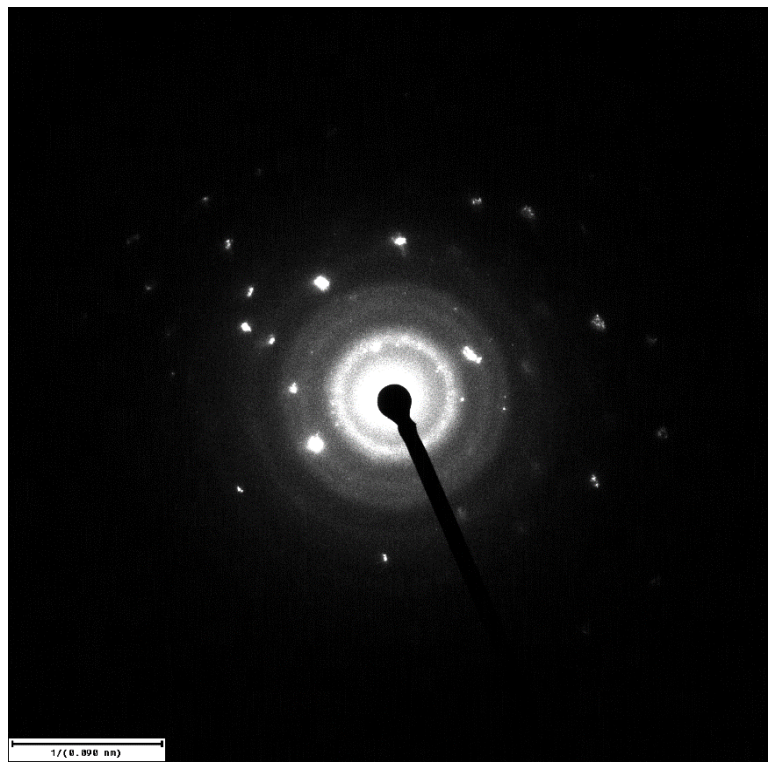

*Figure S5. Selected area electron diffraction pattern from the oxide region in Figure 1a*

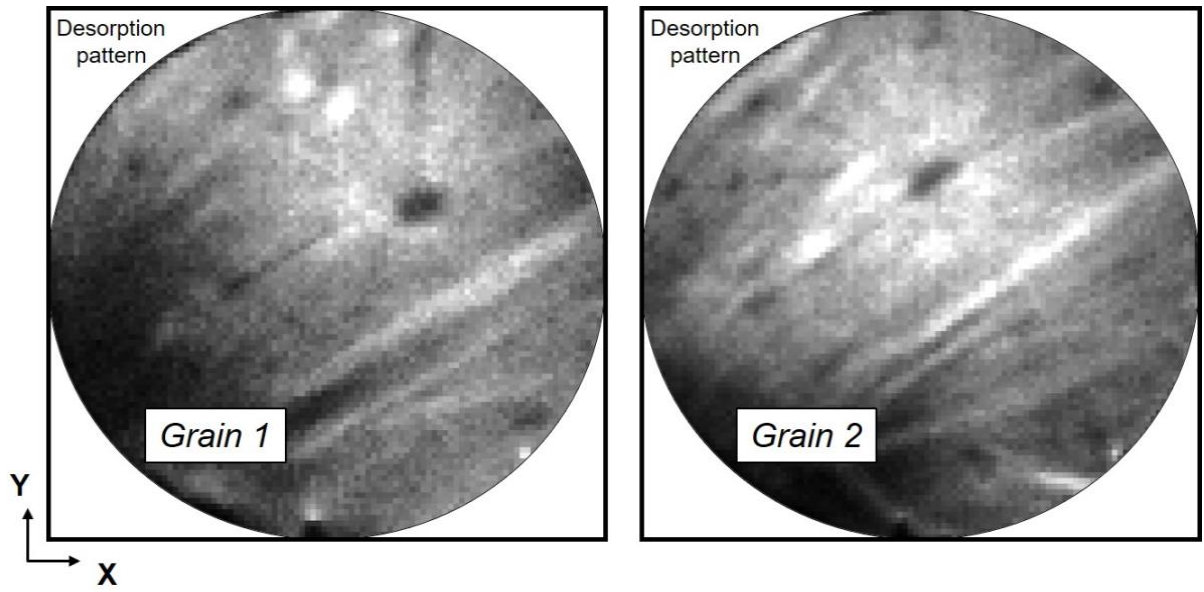

*Figure S6. Desorption pattern showing two distinct crystallographic orientations within the crack tip (Figure 2b)*

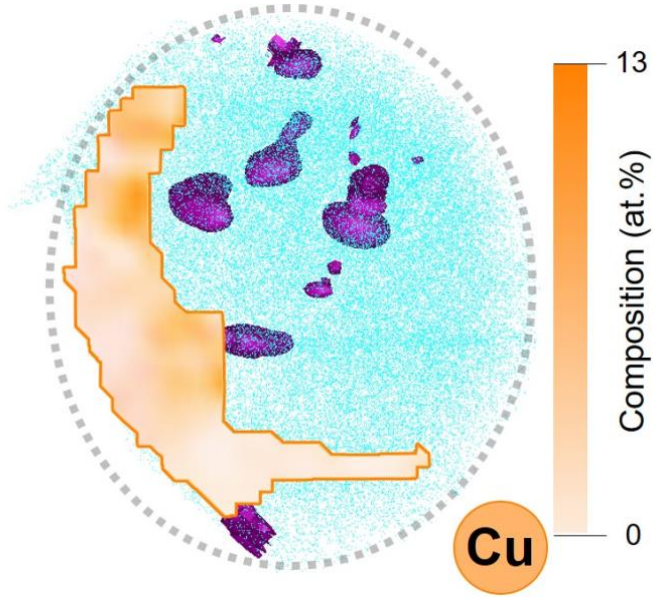

*Figure S7. 2D Cu composition map, obtained from a 10 nm slice, showing the 2D elemental distribution of Cu in the oxide within the dotted-line slice shown in Figure 2b.*